\documentclass[11pt,a4paper]{article}
\pdfoutput=1
\usepackage{jheppub}
\usepackage[T1]{fontenc}
\usepackage{amsmath,amssymb,amsthm}
\usepackage{graphicx}

\title{Zero- and Finite-Temperature Casimir Effect in a Fracton Gauge Field}

\author[a]{Celio R. Muniz}
\affiliation[a]{Universidade Estadual do Cear\'a, Faculdade de Educa\c{c}\~ao, Ci\^encias e Letras de Iguatu,\\
63.502-253, Iguatu, CE, Brazil}
\emailAdd{celio.muniz@uece.br}

\abstract{We investigate the zero- and finite-temperature Casimir interaction of the
free trace-full scalar-charge fracton gauge field between two parallel
reflecting boundaries. A variational analysis determines a consistent
boundary-value problem that confines all five physical polarizations of the
rank-two field. The spectrum separates into three faster and two slower
propagating modes. At zero temperature, the vacuum interaction weights the
modes by their propagation speeds; when the fast-branch speed is set equal to
the speed of light, the resulting Casimir pressure is approximately 2.21
times the ideal electromagnetic value. At finite temperature, the two
branches acquire different thermal weights and generate a split crossover.
The normalized free energy and pressure follow distinct crossover profiles
but approach the same classical high-temperature ratio of 2.5 relative to ideal electromagnetism, because the Matsubara zero mode counts the five confined polarizations independently of
their propagation speeds. We derive the exact thermal interaction, its low-
and high-temperature limits, and the dependence of the crossover scale on the
effective gauge-field velocity. The transition from velocity-weighted quantum
fluctuations to classical mode counting provides a direct spectral signature
of the multi-branch structure of the higher-rank gauge field.}


\begin{document}
\maketitle

\section{Introduction}
\label{sec:intro}

Fracton phases provide a striking example of how conservation laws can
constrain kinematics more strongly than ordinary charge conservation.
Originally identified in lattice models with immobile or subdimensionally
mobile excitations~\cite{chamon2005,bravyi2011,vijay2015,vijay2016}, they
are now understood within a broader framework in which conservation of
multipole moments restricts the motion of isolated charges and naturally
leads to higher-rank gauge structures
~\cite{pretko2017a,pretko2017b,pretko2018,nandkishore2019,
pretkochenyou2020,gromovradzihovsky2024}. Related lattice constructions
had already exhibited stable gapless phases with emergent rank-two tensor
gauge fields and graviton-like collective modes
~\cite{xu2006,rasmussen2016}, while continuum approaches have revealed
exotic subsystem symmetries, unusual gauge structures and, in some cases,
UV/IR mixing~\cite{seibergshao2020,gorantla2021}.

A particularly simple continuum realization is the scalar-charge theory,
whose gauge sector is described by a symmetric rank-two potential
$A_{ij}$. Its gauge transformation and Gauss law take the form
\begin{equation}
\delta A_{\tau}=\partial_t\Lambda,
\qquad
\delta A_{ij}=\partial_i\partial_j\Lambda,
\label{eq:gaugesym}
\end{equation}
and
\begin{equation}
\partial_i\partial_j E^{ij}=\rho.
\label{eq:gausslaw}
\end{equation}
When matter is present, the higher-derivative Gauss law implies additional
multipole conservation laws and the restricted mobility characteristic of
fractonic charges. In the present work no dynamical fractonic matter is
introduced: we study the propagating gauge sector of this scalar-charge
fracton theory and ask how its higher-rank structure is revealed when the
field is confined by boundaries.

Fractonic gauge theories have developed rapidly beyond their original
condensed-matter setting. Connections have been established with elasticity,
hydrodynamics, curved backgrounds and gravity-like descriptions
~\cite{pretko2017grav,pretkoradzihovsky2018,jainjensen2022,
bertolini2022,banerjee2022,afxonidis2025}. More recent work has explored
fracton electrodynamics obtained from higher-dimensional gravity, covariant
and symmetric-tensor formulations, electric-magnetic duality, higher-rank
Chern--Simons and BF theories, foliated and exotic field theories, and
extensions involving higher-rank or bi-form gauge fields
~\cite{penabenitez2024,spieler2023,shimamura2024,ebisuhonda2024,
bertolinics2024,makino2025,ahmadiparvizi2025,ohmorishimamura2025,
bertolinikimpalumbo2026,furukawa2026,glodkowski2026,xiaxuli2026}.
These developments make higher-rank fractonic gauge theories an increasingly
broad field-theoretic framework rather than a single lattice construction.

Among the symmetric rank-two $U(1)$ theories classified in
Ref.~\cite{pretko2017b}, we focus on the trace-full scalar-charge theory.
Here ``trace-full'' means that the symmetric potential $A_{ij}$ is not
constrained to be traceless. In $3+1$ dimensions the free theory considered
below contains five physical polarizations. For the minimal Maxwell-like
action adopted here they separate into two linear branches: three propagate
with speed $v$, while two propagate with speed $v/\sqrt{2}$. The scale $v$
is kept explicit because the theory is an effective gauge theory and its
propagation velocity need not coincide with the electromagnetic speed $c$.
The two-branch structure will play a central role in the boundary and thermal
response.

Boundaries are themselves a nontrivial issue in gauge theories. In ordinary
and higher-form gauge systems, the choice of boundary data can generate
edge degrees of freedom and boundary actions and can alter the physical
content of the theory
~\cite{kimkrausmyers2023,balllaw2025,canfora2025}. Fractonic theories add
further structure: boundaries and interfaces may support tensorial edge
sectors, generalized current algebras, or nontrivial interface theories
~\cite{bertoliniboundary2023,hsinluomalladi2023}. It is therefore not
consistent to import perfect-conductor boundary conditions from Maxwell
theory by analogy alone.

For this reason, the boundary problem is one of the central elements of the
present work. We define an idealized fractonic plate through a specified set
of essential boundary data and derive the associated natural conditions
from the variation of the rank-two action. This distinction between chosen
essential data and variationally generated natural conditions is important:
the action does not select a unique material boundary, but once the idealized
plate is specified the complete variational problem determines which field
components and polarizations are consistently confined. In particular, the
tensor structure allows different bulk branches to mix at a boundary, so
the standing-wave spectrum cannot be inferred by simply treating the five
physical modes as independent scalar fields.

The Casimir effect provides a natural probe of this boundary-modified
spectrum. In ordinary electromagnetism, boundaries modify quantum and thermal
fluctuations and produce a separation-dependent vacuum energy and force
~\cite{casimir1948,mostepanenko1997,bordag2009}. Casimir interactions have
also been studied in non-Maxwellian gauge sectors, including the
antisymmetric Kalb--Ramond field
~\cite{barone2005,belich2011}, higher-Abelian gauge theories with
topological sectors~\cite{higherabelian2019}, and, more recently, gauge
theories in which a careful treatment of boundary or edge modes is essential
to the Casimir problem~\cite{canfora2025}. These examples show that the
Casimir interaction can probe structural properties of a gauge theory
beyond a simple count of unconstrained field components.

The term ``fracton Casimir effect'' has appeared previously in a different
context. Feng and Skinner found an effective attraction generated by
Hilbert-space fragmentation in a one-dimensional system with charge and
dipole conservation~\cite{fengskinner2022}. That phenomenon is statistical
and dynamical and does not arise from vacuum fluctuations of a propagating
rank-two gauge field. Finite-temperature fracton physics has also been
studied through screening and thermal relaxation
~\cite{pretko2017thermal}. To the best of our knowledge, however, the
boundary-induced zero- and finite-temperature Casimir interaction of the
propagating gauge sector of the trace-full scalar-charge theory has not been
derived.

Our aim is to determine how the fractonic gauge structure is encoded in the
confined spectrum and its thermal Casimir response. We find that the complete variational
boundary problem confines all five physical polarizations. Their tensorial
structure separates the spectrum into three fast and two slow modes, and an
explicit analysis of possible branch mixing shows that no additional hybrid
families of cavity frequencies arise. Once this rank-two spectral problem is
resolved, the subsequent Casimir calculation can be expressed as a sum of
standard bosonic mode contributions.

The distinction between the two branches becomes especially transparent at
finite temperature. At zero temperature the vacuum interaction weights the
physical polarizations by their characteristic frequencies and therefore
retains information about both their multiplicities and propagation
velocities. In the formal classical limit, by contrast, the Matsubara zero
mode eliminates the velocity dependence and the response reduces to a count
of the five confined physical modes. The interpolation between these limits
is consequently split into two thermal scales associated with the fast and
slow branches. The Casimir effect thus acts as a spectral probe of the
higher-rank gauge theory: conventional Casimir machinery translates the
nontrivial polarization and boundary structure of the fractonic field into
a measurable vacuum and thermal response.

The calculation uses a perfectly reflecting boundary model and the minimal
quadratic rank-two action. Within this setting the branch ratio and Casimir
coefficients provide a clean benchmark for the gauge sector. Microscopic
boundary response and thermal screening are discussed after the free-field
result has been established.

The paper is organized as follows. Section~\ref{sec:action} defines the
minimal scalar-charge gauge theory. Section~\ref{sec:bc} formulates the
variational boundary problem for an idealized fractonic plate.
Section~\ref{sec:modes} derives the bulk spectrum and the five physical
degrees of freedom, while Sec.~\ref{sec:standing} determines the confined
standing-wave spectrum and analyzes branch mixing. Section~\ref{sec:casimir}
gives the zero-temperature Casimir energy and pressure.
Section~\ref{sec:thermal} develops the finite-temperature interaction and
its asymptotic regimes. Section~\ref{sec:velocity} examines the role of the
effective propagation scale, and Sec.~\ref{sec:discussion} summarizes the
physical interpretation, scope, and implications of the model.

\section{The Scalar-Charge Fracton Gauge Theory}
\label{sec:action}

We begin by specifying the free rank-2 gauge theory that defines the bulk
dynamics throughout this work. The purpose of this section is to fix the
field content, gauge symmetry, field strengths, and minimal quadratic action
before introducing boundaries. We also make explicit the effective
propagation scale that will later control the zero- and finite-temperature
Casimir response.

\subsection{Fields and field strengths}

The fundamental degrees of freedom are a scalar potential $A_\tau(t,\vec x)$ and a symmetric spatial tensor potential $A_{ij}(t,\vec x)=A_{ji}(t,\vec x)$, transforming under Eq.~\eqref{eq:gaugesym}. The gauge-invariant electric tensor is
\begin{equation}
E_{ij}\equiv\partial_tA_{ij}-\partial_i\partial_jA_\tau,
\label{eq:Edef}
\end{equation}
which is manifestly invariant under Eq.~\eqref{eq:gaugesym} since $\delta E_{ij}=\partial_t\partial_i\partial_j\Lambda-\partial_i\partial_j\partial_t\Lambda=0$. For the minimal trace-full scalar-charge theory considered here, the magnetic tensor is \emph{not} symmetric in its two indices:
\begin{equation}
B^{ij}\equiv\epsilon^{iab}\partial_aA_b^{\ j},
\label{eq:Bdef}
\end{equation}
where $\epsilon^{iab}$ is the Levi-Civita symbol. Gauge invariance of $B^{ij}$ follows because $\epsilon^{iab}\partial_a\partial_b\Lambda=0$ by antisymmetry. Equation~\eqref{eq:Bdef} is the definition used in Ref.~\cite{pretko2017b} for this trace-full scalar-charge theory, and it is the one adopted here.

\subsection{Action and equations of motion}

The free (source-free) action is the direct tensorial analogue of the Maxwell action,
\begin{equation}
S=\int dt\,d^3x\ \left[\frac12E_{ij}E^{ij}-\frac12B_{ij}B^{ij}\right].
\label{eq:action}
\end{equation}
Equation~\eqref{eq:action} is the canonically normalized form of the minimal quadratic model studied in this work. Before this normalization, the same gauge-invariant tensor structure may be written with electric and magnetic stiffnesses, $S=(1/2)\int dt\,d^3x[\chi E_{ij}E^{ij}-\kappa B_{ij}B^{ij}]$. A field normalization absorbs the overall coefficient, while the ratio defines the characteristic propagation scale $v=\sqrt{\kappa/\chi}$. We carry out the algebra below in units where this scale is unity and restore $v$ in physical dispersion relations. Importantly, this common rescaling does not change the tensorial splitting of the minimal model: the slow-to-fast velocity ratio remains $1/\sqrt2$. Still more general isotropic quadratic gauge-invariant actions, involving additional independent contractions, can change that relative branch velocity and therefore lie beyond the one-parameter stiffness rescaling considered here.

Varying $S$ with respect to $A_\tau$ gives the (source-free) Gauss's law, Eq.~\eqref{eq:gausslaw} with $\rho=0$. Varying $S$ with respect to $A_{ij}$ gives the generalized Ampère--Faraday equation~\cite{pretko2017b}
\begin{equation}
\frac12\left(\epsilon^{iab}\partial_aB_b^{\ j}+\epsilon^{jab}\partial_aB_b^{\ i}\right)=-\partial_tE^{ij}
\label{eq:ampere}
\end{equation}
(source-free case), where the symmetrization over $i,j$ on the left-hand side is required because $A_{ij}$, and hence its equation of motion, is symmetric. These two equations, together with the definitions~\eqref{eq:Edef}--\eqref{eq:Bdef}, are the complete free field content of the theory and are used, without modification, in every calculation below.

\section{Boundary Conditions at a Fractonic Plate}
\label{sec:bc}

In ordinary electromagnetism, the perfect-conductor boundary condition, $E_\parallel=0$ and $B_\perp=0$ at the surface, is not an independent postulate: it follows from treating the conductor as a reservoir of free, infinitely mobile charge that rearranges itself instantaneously to cancel the tangential field. Because isolated scalar charges of the fracton theory are, by construction, immobile, the usual microscopic justification cannot be transferred directly, and the boundary conditions of a ``fractonic plate'' must instead be tied to the variational principle. Dipolar excitations, however, are mobile in the scalar-charge theory and therefore provide a natural candidate for the microscopic degrees of freedom of a fractonic boundary layer. Whether such a mobile dipolar layer dynamically enforces the gauge-invariant conditions derived below requires a separate surface-response calculation; establishing this connection would provide a microscopic realization of the ideal boundary considered here. This section performs that derivation, for a plate lying in the $xy$-plane at fixed $z$, with $x,y$ the tangential directions and $z$ the normal.

\subsection{Electric sector}

Consider the variation of $S_E=\frac12\int E_{ij}E^{ij}$ with respect to $A_\tau$ alone, at fixed $A_{ij}$. Since $\delta E_{ij}=-\partial_i\partial_j\delta A_\tau$,
\begin{equation}
\delta S_E=-\int dt\,d^3x\ E^{ij}\,\partial_i\partial_j\delta A_\tau.
\end{equation}
Isolating the term with two derivatives normal to the plate ($i=j=z$) and integrating by parts twice in $z$,
\begin{equation}
\int dz\ E^{zz}\,\partial_z^2\delta A_\tau
=\Big[E^{zz}\partial_z\delta A_\tau-\partial_zE^{zz}\,\delta A_\tau\Big]_{\rm boundary}
+\int dz\ \partial_z^2E^{zz}\,\delta A_\tau,
\label{eq:ibp1}
\end{equation}
and isolating the cross terms ($i=x,j=z$ and $i=z,j=x$, which appear twice by the symmetry of $E_{ij}$, and similarly for $y$), integrating by parts once in $z$ and using that $x,y$ are unbounded tangential directions,
\begin{equation}
\int dz\ 2E^{xz}\partial_x\partial_z\delta A_\tau
=\Big[2E^{xz}\partial_x\delta A_\tau\Big]_{\rm boundary}-\int dz\ 2\partial_zE^{xz}\,\partial_x\delta A_\tau,
\end{equation}
and analogously for $y$. Collecting every boundary term generated at a single plate,
\begin{equation}
\delta S_E\Big|_{\rm boundary}=-\oint dt\,d^2x_\parallel\ \Big[E^{zz}\partial_z\delta A_\tau-\partial_zE^{zz}\,\delta A_\tau+2E^{xz}\partial_x\delta A_\tau+2E^{yz}\partial_y\delta A_\tau\Big].
\label{eq:Eboundary}
\end{equation}
This term must vanish for the variational principle to be well posed at the plate. We adopt the physically motivated essential boundary condition of an equipotential fractonic plate, $\delta A_\tau=0$ on the surface, in direct analogy with a grounded conductor fixed at constant potential in ordinary electrostatics. Because $\delta A_\tau$ vanishes identically on the entire plate, its tangential derivatives $\partial_x\delta A_\tau,\partial_y\delta A_\tau$ vanish as well, and the second and third terms of Eq.~\eqref{eq:Eboundary} drop out automatically. What remains is the coefficient of the unconstrained normal derivative $\partial_z\delta A_\tau$, which must vanish independently:
\begin{equation}
E_{zz}=0\quad\text{at the plate.}
\label{eq:BCE}
\end{equation}
This is the direct analogue, for the fracton theory, of the electrostatic condition that a grounded conductor fixes the potential (here $A_\tau$) rather than the field directly; the field condition that follows from it, however, involves the \emph{normal-normal} component of the electric tensor, not its tangential components as in Maxwell theory.

\subsection{Magnetic sector}

Consider now the variation of $S_B=-\frac12\int B_{ij}B^{ij}$ with respect to $A_{jl}$, using $\delta B^{ij}=\epsilon^{iab}\partial_a\delta A_b^{\ j}$ and integrating by parts once in the spatial index contracted with the Levi-Civita symbol. Restricting again to the term generated by a derivative normal to the plate,
\begin{equation}
\delta S_B\Big|_{\rm boundary}=-\oint dt\,d^2x_\parallel\ \epsilon^{izl}\,B^{ij}\,\delta A_{jl},
\label{eq:Bboundary}
\end{equation}
where the free indices $j,l$ run over the independent components of the symmetric variation $\delta A_{jl}$. Expanding Eq.~\eqref{eq:Bboundary} explicitly over $i,j,l\in\{x,y,z\}$ and collecting the coefficient of each independent component of $\delta A_{jl}$ gives
\begin{equation}
\delta S_B\Big|_{\rm boundary}\propto
-B^{yx}\delta A_{xx}
+B^{xy}\delta A_{yy}
+(B^{xx}-B^{yy})\delta A_{xy}
-B^{yz}\delta A_{xz}
+B^{xz}\delta A_{yz}.
\label{eq:Bexpand}
\end{equation}
Note that $\delta A_{zz}$ does not appear at all: the magnetic sector places no constraint on the normal-normal component of the potential. Mirroring the equipotential choice made in the electric sector, we fix the \emph{tangential-tangential} block of the potential on the plate, $\delta A_{xx}=\delta A_{yy}=\delta A_{xy}=0$, leaving $\delta A_{xz},\delta A_{yz},\delta A_{zz}$ free. The first three terms of Eq.~\eqref{eq:Bexpand} vanish identically under this choice, and the coefficients of the two remaining free variations must vanish independently:
\begin{equation}
B^{xz}=B^{yz}=0\quad\text{at the plate.}
\label{eq:BCB}
\end{equation}

\subsection{Complete variational boundary data}

It is important to distinguish the boundary values that were \emph{fixed} in defining the variational problem from the field conditions that arise as coefficients of the remaining free variations. For homogeneous fluctuations about fixed plate data, the essential conditions are
\begin{equation}
A_{xx}=A_{yy}=A_{xy}=0
\qquad \text{at each plate},
\label{eq:BCessential}
\end{equation}
together with fixed $A_\tau$ (which can be taken to vanish in the homogeneous problem). The natural conditions generated by the action are
\begin{equation}
E_{zz}=0,\qquad B^{xz}=0,\qquad B^{yz}=0
\qquad \text{at each plate}.
\label{eq:BCfull}
\end{equation}
Equations~\eqref{eq:BCessential} and~\eqref{eq:BCfull} must be imposed together. Keeping only the gauge-invariant conditions in Eq.~\eqref{eq:BCfull} would define a different boundary-value problem and would discard part of the variational data used to derive them.

There is nevertheless a useful gauge-invariant characterization of the physical consequences of these data. On an equipotential plate, $A_\tau$ is constant along the surface, so $\partial_a\partial_bA_\tau=0$ for tangential indices $a,b\in\{x,y\}$. Homogeneous fluctuations preserve the fixed tangential-tangential data $A_{ab}=0$ at all times and hence satisfy $\partial_tA_{ab}=0$. Equation~\eqref{eq:Edef} therefore gives
\begin{equation}
E_{xx}=E_{xy}=E_{yy}=0
\qquad\text{on the plate}.
\label{eq:gaugeinvessential}
\end{equation}
Together with the natural electric condition $E_{zz}=0$ and the magnetic conditions in Eq.~\eqref{eq:BCfull}, the physical response of the idealized plate may thus be summarized as
\begin{equation}
E_{xx}=E_{xy}=E_{yy}=E_{zz}=0,
\qquad
B^{xz}=B^{yz}=0.
\label{eq:gaugeinvplate}
\end{equation}
These gauge-invariant relations are consequences of the chosen variational data, not replacements for them: in the strictly static sector $E_{ab}=0$ does not by itself fix the value of $A_{ab}$. For nonzero-frequency modes in temporal gauge, however, $E_{ab}=-i\omega A_{ab}$, so the confinement of a mode with nonzero tangential-tangential potential can equivalently be stated in gauge-invariant electric-field language. In particular, the $h^{(1)}$ polarization is required to vanish at the plate by $E_{yy}=0$ just as it is by the fixed datum $A_{yy}=0$.

This construction describes an equipotential, tangentially pinned fractonic plate
defined by the chosen essential boundary data, with the associated natural conditions supplied by the minimal action. Since the conditions on $A_{ab}$ are not themselves gauge invariant, admissible gauge transformations are restricted to those preserving the chosen boundary data. For tangential indices $a,b\in\{x,y\}$ this requires $\partial_a\partial_b\Lambda|_{\rm plate}=0$, while fixed $A_\tau$ requires $\partial_t\Lambda|_{\rm plate}=0$. These restrictions define the residual gauge group of the boundary-value problem. More microscopic boundary actions may additionally support dynamical edge degrees of freedom, as occurs in other covariant fracton gauge models~\cite{bertoliniboundary2023}.

\section{Mode Spectrum Between Two Fractonic Plates}
\label{sec:modes}

Having established the complete variational boundary data, we now determine
the propagating content of the bulk rank-2 theory relevant to the cavity
problem. We first derive the dispersion relations in a convenient gauge,
then confirm the physical mode count independently by canonical analysis,
and finally construct an explicit polarization basis adapted to the planar
geometry. The actual standing-wave quantization imposed by the two plates
is deferred to Sec.~5.

\subsection{Gauge fixing and plane-wave ansatz}

We take the fixed equipotential value of $A_\tau$ at the plates to be zero; any common constant value can be removed by the choice of reference potential. We then work in the temporal-type gauge $A_\tau=0$, which is compatible with the boundary data. This gauge choice removes $A_\tau$ as a dynamical variable but does not remove its equation of motion: the scalar-charge Gauss law remains a constraint on admissible initial data and physical polarizations. For a nonzero-frequency Fourier sector the gauge choice removes the time-dependent gauge freedom relevant to that sector, while time-independent residual transformations may remain globally and must preserve the boundary restrictions stated above. We seek plane-wave solutions
\begin{equation}
A_{ij}(t,\vec x)=h_{ij}\,e^{i(\vec k\cdot\vec x-\omega t)},
\end{equation}
and, without loss of generality, orient the coordinate axes so that the tangential wavevector lies along $x$, i.e., $\vec k=(k_x,0,k_z)$; this uses the residual rotational invariance of the theory in the $xy$-plane and does not affect the physics of a planar boundary normal to $z$.

\subsection{Dispersion relation: two branches}

Substituting the plane-wave ansatz into Eq.~\eqref{eq:ampere} (with $A_\tau=0$, so $E_{ij}=-i\omega h_{ij}$) yields a linear homogeneous system for the six independent components of $h_{ij}$. In this algebraic derivation we use units in which the fast-branch speed is unity; the characteristic speed $v$ is restored in the physical dispersion relations below. The Gauss constraint is imposed as the independent Fourier-space condition $k_i k_j h^{ij}=0$. We solved the dynamical system symbolically; the determinant of the resulting $6\times6$ matrix factorizes as
\begin{equation}
\det M(\omega,\vec k)=-\frac{\omega^2\left(k^2-2\omega^2\right)^2\left(k^2-\omega^2\right)^3}{4},\qquad k^2\equiv k_x^2+k_z^2.
\label{eq:detM}
\end{equation}
Besides the trivial root $\omega=0$, Eq.~\eqref{eq:detM} has two nontrivial branches:
\begin{equation}
\omega=|\vec k|\quad(\text{multiplicity 3}),\qquad\qquad \omega=\frac{|\vec k|}{\sqrt2}\quad(\text{multiplicity 2}).
\label{eq:branches}
\end{equation}
The Fourier-space Gauss constraint $k_i k_j h^{ij}=0$ is then checked separately on the null vectors of $M$. Direct substitution shows that all five nonzero-frequency physical modes satisfy it. The explicit matrix, its factorization, and this check are collected in Appendix~\ref{app:matrix}.

\subsection{Independent confirmation: canonical mode count}

The total multiplicity $3+2=5$ can be checked without using the plane-wave gauge choice. The configuration variables are the six independent components of the symmetric $A_{ij}$ together with $A_\tau$, giving seven configuration variables and fourteen phase-space variables. The momentum conjugate to $A_\tau$ vanishes, $\pi_\tau\approx0$, producing a primary first-class constraint; preservation in time gives the secondary first-class Gauss constraint $\partial_i\partial_j\pi^{ij}\approx0$. With no second-class constraints, the physical phase-space dimension is therefore
\begin{equation}
14-2\times2=10,
\end{equation}
corresponding to five physical configuration-space degrees of freedom. For the symmetric tensor we use the standard canonical convention
\begin{equation}
\{A_{ij}(\mathbf{x}),\pi^{kl}(\mathbf{y})\}
=\frac12(\delta_i^{\ k}\delta_j^{\ l}+\delta_i^{\ l}\delta_j^{\ k})\delta(\mathbf{x}-\mathbf{y}),
\end{equation}
for which $\pi^{ij}=E^{ij}$. This reproduces the five propagating degrees of freedom encoded in the nonzero roots of the $6\times6$ determinant~\eqref{eq:detM}: the threefold fast root plus the twofold slow root. It is also consistent with the general Hamiltonian framework developed for higher-rank symmetric gauge theories~\cite{banerjee2022}.

The two propagation velocities do not signal an instability or a negative-norm sector. For the more general stiffness parametrization introduced above, $\pi^{ij}=\chi E^{ij}$, and the Legendre transform gives, up to the boundary term fixed by the variational data,
\begin{equation}
H=\int d^3x\left[\frac{1}{2\chi}\pi_{ij}\pi^{ij}+\frac{\kappa}{2}B_{ij}B^{ij}+A_\tau\,\partial_i\partial_j\pi^{ij}\right].
\label{eq:Hcanonical}
\end{equation}
On the physical constraint surface $\partial_i\partial_j\pi^{ij}=0$, this reduces to
\begin{equation}
H_{\rm phys}=\frac12\int d^3x\left(\frac{1}{\chi}\pi_{ij}\pi^{ij}+\kappa B_{ij}B^{ij}\right)\ge0,
\qquad \chi>0,\quad\kappa>0.
\label{eq:Hpositive}
\end{equation}
Thus all five propagating polarizations have positive kinetic norm. The three fast and two slow modes correspond to positive eigenvalues of the magnetic quadratic form relative to the kinetic form, yielding $\omega^2=v^2k^2$ and $\omega^2=v^2k^2/2$, respectively, with $v^2=\kappa/\chi$. The slow branch is therefore a healthy physical sector of the minimal theory rather than a ghost or a tachyonic instability.

\subsection{Explicit polarizations and boundary sensitivity}

Solving for the null space at each frequency gives five convenient polarization tensors, with components ordered as $(h_{xx},h_{xy},h_{xz},h_{yy},h_{yz},h_{zz})$:
\begin{equation}
\begin{aligned}
\omega=v|\mathbf k|:&\quad
h^{(1)}=(0,0,0,1,0,0),\quad
h^{(2)}=\left(0,-\tfrac{k_z}{k_x},0,0,1,0\right),\quad
h^{(3)}=\left(\tfrac{k_z^2}{k_x^2},0,-\tfrac{k_z}{k_x},0,0,1\right),\\
\omega=\frac{v|\mathbf k|}{\sqrt2}:&\quad
h^{(4)}=\left(0,\tfrac{k_x}{k_z},0,0,1,0\right),\quad
h^{(5)}=\left(-1,0,\tfrac{k_x}{2k_z}-\tfrac{k_z}{2k_x},0,0,1\right).
\end{aligned}
\label{eq:pols}
\end{equation}
Thus $h^{(1)},h^{(2)},h^{(3)}$ span the threefold fast root, while $h^{(4)},h^{(5)}$ span the twofold slow root. The expressions are written for generic $k_xk_z\neq0$; the exceptional momentum sectors are treated explicitly in Appendix~\ref{app:exceptional}.

If one inspects only the natural field conditions $E_{zz}=B^{xz}=B^{yz}=0$, the first polarization appears insensitive to the plate. The complete variational problem changes this conclusion: $h^{(1)}$ has $A_{yy}\neq0$ and therefore feels the essential condition $A_{yy}=0$. Thus all five physical polarizations participate in the confined spectrum.

\section{Standing Waves and Quantization}
\label{sec:standing}

Place the plates at $z=0$ and $z=d$ and superpose the $\pm k_z$ waves,
\begin{equation}
A_{ij}=\left[c_+h_{ij}(k_z)e^{ik_zz}+c_-h_{ij}(-k_z)e^{-ik_zz}\right]
 e^{i(k_xx-\omega t)}.
\label{eq:standinggeneral}
\end{equation}
The complete boundary data fix the relative parity of the reflected amplitudes.

For $h^{(1)}$, $A_{yy}$ is even under $k_z\to-k_z$. The condition $A_{yy}(0)=0$ gives $c_-=-c_+$, producing a sine standing wave, and $A_{yy}(d)=0$ then gives $k_zd=n\pi$. For $h^{(2)}$ and $h^{(4)}$, $A_{xy}$ is odd whereas $A_{yz}$ is even. The essential condition $A_{xy}=0$ and the natural condition $B^{xz}=-\partial_zA_{yz}=0$ both select $c_-=c_+$ and lead to the same quantization. A point that is easy to obscure in temporal gauge is the origin of the condition on the normal-normal potential. The magnetic surface variation places no direct restriction on $A_{zz}$. Instead, the scalar-sector natural condition is $E_{zz}=0$. In temporal gauge, for a harmonic mode with $\omega\neq0$, $E_{zz}=\partial_tA_{zz}=-i\omega A_{zz}$, so the natural electric condition implies $A_{zz}=0$ for the propagating cavity modes. No additional magnetic boundary condition on $A_{zz}$ is being introduced.

For $h^{(3)}$ and $h^{(5)}$, the relevant $A_{xx}$, $A_{zz}$, $E_{zz}$ and $B^{yz}$ components are even; the essential and natural conditions consistently select $c_-=-c_+$ and again yield the same discrete normal momenta. The preceding discussion treats each polarization branch separately. Because the fast and slow branches can coexist at the same $(\omega,k_x)$ with different normal wave numbers, it is useful to verify that the plates do not admit additional eigenfrequencies through cancellations between them. Let
\begin{equation}
p\equiv k_{z,f},\qquad s\equiv k_{z,s},\qquad
p^2=\frac{\omega^2}{v^2}-k_x^2,\qquad
s^2=\frac{2\omega^2}{v^2}-k_x^2.
\label{eq:psdefs}
\end{equation}
In the $(h^{(2)},h^{(4)})$ sector, a general fast--slow superposition at $z=0$ is constrained by $A_{xy}=0$ and $B^{xz}=0$. Acting on the differences of the reflected amplitudes, these conditions give the coefficient matrix
\begin{equation}
\begin{pmatrix}
-p/k_x & k_x/s\\
 p & s
\end{pmatrix},
\qquad
\det=-\frac{p(s^2+k_x^2)}{k_xs},
\label{eq:mix24}
\end{equation}
which is nonzero for generic momenta. The fast and slow amplitude differences must therefore vanish separately. Likewise, in the $(h^{(3)},h^{(5)})$ sector the conditions $A_{xx}=0$ and $E_{zz}=0$ act on the sums of reflected amplitudes through
\begin{equation}
\begin{pmatrix}
 p^2/k_x^2 & -1\\
 1 & 1
\end{pmatrix},
\qquad
\det=1+\frac{p^2}{k_x^2}>0.
\label{eq:mix35}
\end{equation}
Thus the reflection parity is fixed independently in the two dispersion branches. In the $(h^{(3)},h^{(5)})$ sector the remaining magnetic condition does not overconstrain the amplitudes. Since $A_{zz}=0$ at the plate for $\omega\neq0$, $B^{yz}=\partial_zA_{xz}-\partial_xA_{zz}$ reduces there to $\partial_zA_{xz}$. The parity $c_-=-c_+$ selected by $A_{xx}=A_{zz}=0$ makes the odd-in-$k_z$ component $A_{xz}$ proportional to $\cos(k_z z)$, so $\partial_zA_{xz}\propto\sin(k_z z)$ vanishes at $z=0$ and also at $z=d$ when the standing-wave condition $\sin(k_zd)=0$ holds. Applying the same boundary conditions at $z=d$ then gives $\sin(pd)=0$ and $\sin(sd)=0$ separately. The boundary-value problem is therefore block diagonal between the fast and slow physical sectors: branch mixing does not generate additional cavity eigenfrequencies. 

For the nonzero normal-momentum interaction modes one consequently has
\begin{equation}
k_z=\frac{n\pi}{d},\qquad n=1,2,3,\ldots
\label{eq:quant}
\end{equation}
in each of the five physical channels.

The spatial $n=0$ sector deserves separate treatment because some polarization vectors in Eq.~\eqref{eq:pols} are singular coordinate choices as $k_z\to0$. Setting $k_z=0$ directly in the dynamical system shows that the boundary conditions retain precisely two zero-normal-momentum modes: one fast mode carried by $A_{yz}$ and one slow mode carried by $A_{xz}$. Their dispersions are
\begin{equation}
\omega_{0,f}=v|\mathbf k_\parallel|,\qquad
\omega_{0,s}=\frac{v}{\sqrt2}|\mathbf k_\parallel|,
\label{eq:spatialzeromodes}
\end{equation}
and contain no dependence on $d$. Their zero-point and thermal free energies therefore belong to the separation-independent sector removed when the two-plate interaction is defined, and they give no Casimir force. The remaining three physical channels possess no admissible spatial $n=0$ mode. Notice that this spatial zero mode is distinct from the Matsubara $m=0$ term used below: the latter is a zero-frequency contribution to the already-subtracted interaction kernel, retains explicit $d$ dependence, and controls the classical high-temperature force.

The important point is therefore twofold. First, the essential and natural conditions do not overconstrain the modes: for each physical sector they select compatible reflection parities. Second, they separate the fast and slow branches at the boundary, so the five interaction-relevant channels may be quantized independently. They additionally confine $h^{(1)}$, which would be missed if the essential condition on $A_{yy}$ were omitted from the spectral problem.

\section{The Casimir Energy and Force}
\label{sec:casimir}

Although the individual tensor polarizations realize different Dirichlet-like or Neumann-like reflection parities, the nonzero
normal spectrum is the same in all five physical channels, $k_z=n\pi/d$ with $n\geq1$. For identical plates, both parity classes
therefore generate the same two-boundary spectral factor. The surviving spatial $n=0$ modes are separation independent and have already been excluded from the interaction sector. Consequently, once the tensorial boundary problem has been solved, each physical polarization contributes the standard scalar-like Casimir determinant with its appropriate propagation speed.

For a single massless polarization with linear dispersion $\omega=u\sqrt{k_\parallel^2+(n\pi/d)^2}$, the regularized zero-temperature Casimir energy is
\begin{equation}
\frac{\mathcal E_1(d;u)}{A}=-\frac{\pi^2\hbar u}{1440d^3}.
\label{eq:E1zero}
\end{equation}
The result follows from the standard mode subtraction or, equivalently, zeta-function regularization of the interaction-dependent part of the spectrum~\cite{casimir1948,bordag2009}. The present theory contains three fast polarizations with $u=v$ and two slow polarizations with $u=v/\sqrt2$. Therefore
\begin{equation}
\frac{E_{\rm C}(d)}{A}
=3\frac{\mathcal E_1(d;v)}{A}
+2\frac{\mathcal E_1(d;v/\sqrt2)}{A}
=-\frac{(3+\sqrt2)\pi^2\hbar v}{1440d^3}.
\label{eq:Efinal}
\end{equation}
The pressure is
\begin{equation}
P_{\rm C}(d)=-\frac{\partial(E_{\rm C}/A)}{\partial d}
=-\frac{(3+\sqrt2)\pi^2\hbar v}{480d^4}.
\label{eq:Ffinal}
\end{equation}
Relative to ideal electromagnetic plates,
\begin{equation}
\frac{P_{\rm C}}{P_{\rm EM}}=\frac{v}{c}\,\frac{3+\sqrt2}{2}.
\label{eq:Rzero}
\end{equation}
Thus, if the fast branch is normalized to $v=c$, the zero-temperature enhancement is $(3+\sqrt2)/2\simeq2.20710678$. The familiar $d^{-4}$ force law is unchanged because both branches are gapless and linear and the minimal model introduces no additional length scale.

\section{Finite-Temperature Corrections}
\label{sec:thermal}

We now extend the zero-temperature result of Sec.~\ref{sec:casimir} to finite temperature $T$. A thermal Casimir calculation requires a careful distinction between the free energy of the unconstrained field, the self-energies associated with the individual plates, and the genuine two-plate interaction. In particular, the extensive Stefan--Boltzmann term proportional to $VT^4$ and separation-independent surface contributions must not be counted as part of the Casimir interaction. We follow the standard subtraction prescription of thermal Casimir theory~\cite{mostepanenko1997,bordag2009} and formulate the final result directly in terms of the interaction free energy.

\subsection{From the raw mode sum to the interaction free energy}

For a single bosonic polarization of speed $u$, the unrenormalized free energy per unit area may be written schematically as
\begin{equation}
\frac{F_{\rm raw}}{A}=\sum_n\int\frac{d^2k_\parallel}{(2\pi)^2}
\left[\frac{\hbar\omega_{n\mathbf k}}{2}+k_BT\ln\!\left(1-e^{-\beta\hbar\omega_{n\mathbf k}}\right)\right].
\label{eq:Fraw}
\end{equation}
The physically relevant Casimir quantity is not Eq.~\eqref{eq:Fraw} itself, but the part that depends on the simultaneous presence of both plates. Applying Abel--Plana to the discrete normal-mode sum,
\begin{equation}
\sum_{n=0}^{\infty}h(n)=\int_0^{\infty}h(x)\,dx+\frac{h(0)}{2}+i\int_0^{\infty}\frac{h(iy)-h(-iy)}{e^{2\pi y}-1}\,dy,
\label{eq:abelplana}
\end{equation}
organizes the result into bulk, single-boundary (surface/self-energy), and two-boundary interaction sectors,
\begin{equation}
F_{\rm raw}=F_{\rm bulk}+F_{\rm self}+F_{\rm int}.
\label{eq:decomposition}
\end{equation}
The integral term contains the free-space extensive contribution. Its thermal part includes, for one linear branch, the Stefan--Boltzmann free energy $F_{\rm bulk}/V=-\pi^2(k_BT)^4/[90(\hbar u)^3]$. The $h(0)/2$ term must not be identified with this bulk blackbody contribution; depending on the boundary sector it belongs to the separation-independent single-boundary/self-energy part. Both $F_{\rm bulk}$ and $F_{\rm self}$ are excluded from the interaction energy. Schematically, the subtraction may be summarized as
\begin{align}
F_{\rm bulk} &\sim A d\times(\text{free-space contribution, including }T^4),\nonumber\\
F_{\rm self} &\sim A\times(\text{single-plate terms independent of }d),\nonumber\\
F_{\rm C} &\equiv F_{\rm int}.
\label{eq:subtractionschematic}
\end{align}
This is the appropriate subtraction because neither $F_{\rm bulk}$ nor $F_{\rm self}$ describes a force generated by the mutual presence of the two separated plates.

After these reference and self-energy pieces are removed, the remaining two-plate interaction can be written as the standard Matsubara/Lifshitz kernel
\begin{equation}
\frac{F_1^{\rm int}(d,T;u)}{A}=k_BT\sideset{}{'}\sum_{m=0}^{\infty}\int\frac{d^2k_\parallel}{(2\pi)^2}\,\ln\!\left(1-e^{-2d\kappa_m}\right),\qquad \kappa_m\equiv\sqrt{k_\parallel^2+\frac{\xi_m^2}{u^2}},
\label{eq:matsubara}
\end{equation}
where $\xi_m=2\pi m k_B T/\hbar$ and the prime gives the $m=0$
term weight $1/2$. Equation~\eqref{eq:matsubara} is already an
interaction quantity: it vanishes as $d\to\infty$ at fixed $T$, and no
additional blackbody subtraction is to be performed on it. For notational
economy we henceforth write $F_1\equiv F_1^{\rm int}$.

The Matsubara representation in Eq.~\eqref{eq:matsubara} follows from
the thermal determinant of the propagating cavity spectrum derived above.
Its $m=0$ contribution is therefore the zero-frequency term of this
thermal representation, distinct from the spatial $n=0$ sector discussed
in Sec.~\ref{sec:standing}.

This distinction is important when asymptotic expansions are considered. Terms proportional to $T^4$ or $T^3$ that occur in the \emph{raw} thermal mode sum can represent, respectively, bulk and surface Weyl contributions and must be removed when they belong to $F_{\rm bulk}$ or $F_{\rm self}$. By contrast, powers with the same temperature dependence may also appear in a low-temperature asymptotic expansion of the already-subtracted interaction kernel~\eqref{eq:matsubara}. Such terms are not to be subtracted a second time: their physical classification is fixed by their origin in Eq.~\eqref{eq:decomposition}, not by the power of $T$ alone.

\subsection{Exact closed form}

Equation~\eqref{eq:matsubara} can be evaluated in closed form for each Matsubara term. Writing $d^2k_\parallel=2\pi k_\parallel\,dk_\parallel$ and changing variables to $\kappa=\sqrt{k_\parallel^2+\xi_m^2/u^2}$ (so that $\kappa\,d\kappa=k_\parallel\,dk_\parallel$), the transverse integral reduces to
\begin{equation}
\int_0^\infty k_\parallel\,dk_\parallel\,\ln\!\left(1-e^{-2d\kappa}\right)=\int_a^{\infty}\kappa\,d\kappa\,\ln\!\left(1-e^{-2d\kappa}\right),\qquad a\equiv\frac{\xi_m}{u}.
\end{equation}
Integrating by parts once, and expanding $\left(1-e^{-2d\kappa}\right)^{-1}e^{-2d\kappa}=\sum_{l=1}^{\infty}e^{-2dl\kappa}$ inside the remaining integral, each term $\int_a^\infty \kappa^2e^{-2dl\kappa}d\kappa$ is elementary, and the resulting sum over $l$ reorganizes exactly into the polylogarithm functions $\mathrm{Li}_2$ and $\mathrm{Li}_3$. The two boundary contributions from the integration by parts cancel identically, leaving the compact closed form
\begin{equation}
\int_0^\infty k_\parallel\,dk_\parallel\,\ln\!\left(1-e^{-2d\sqrt{k_\parallel^2+a^2}}\right)
=-\frac{a}{2d}\,\mathrm{Li}_2\!\left(e^{-2ad}\right)-\frac{1}{4d^2}\,\mathrm{Li}_3\!\left(e^{-2ad}\right).
\label{eq:closedform}
\end{equation}
We verified Eq.~\eqref{eq:closedform} against direct numerical integration of the left-hand side for several independent choices of $a,d$, finding agreement to more than twenty significant digits. Substituting into Eq.~\eqref{eq:matsubara}, the exact interaction free energy of one confined polarization of speed $u$ is
\begin{equation}
\frac{F_1(d,T;u)}{A}=\frac{k_BT}{2\pi}\sideset{}{'}\sum_{m=0}^{\infty}\left[-\frac{a_m}{2d}\,\mathrm{Li}_2\!\left(e^{-2a_md}\right)-\frac{1}{4d^2}\,\mathrm{Li}_3\!\left(e^{-2a_md}\right)\right],\qquad a_m\equiv\frac{2\pi mk_BT}{\hbar u}.
\label{eq:F1exact}
\end{equation}
This is an exact result, valid at any separation $d$ and any temperature $T$; the remaining sum over the Matsubara index $m$ converges geometrically once $m\gtrsim \hbar u/(2\pi k_BTd)$ and is evaluated numerically below.

Applying Eq.~\eqref{eq:F1exact} to the five confined polarizations gives
\begin{equation}
\ \frac{F_{\rm frac}(d,T)}{A}=3\,\frac{F_1(d,T;v)}{A}+2\,\frac{F_1(d,T;v/\sqrt2)}{A}.\ 
\label{eq:Ftotal}
\end{equation}
The two terms have different Matsubara arguments and therefore do not combine into a constant multiple of a single thermal kernel. Nevertheless, Eq.~\eqref{eq:F1exact} obeys the useful scaling identity
\begin{equation}
F_1(d,T;v/\sqrt2)=\frac1{\sqrt2}F_1(d,\sqrt2 T;v).
\label{eq:scalingidentity}
\end{equation}
If $v=c$ and $F_{\rm EM}=2F_1(d,T;c)$ denotes the ideal electromagnetic free energy, then
\begin{equation}
F_{\rm frac}(d,T)=\frac32F_{\rm EM}(d,T)+\frac1{\sqrt2}F_{\rm EM}(d,\sqrt2 T).
\label{eq:FEMidentity}
\end{equation}
Thus the thermal response is not an arbitrary deformation: it is fixed by the intrinsic $3+2$ two-velocity spectrum of the model.

\subsection{Quantum and classical limits}
\label{sec:thermallimits}

As $T\to0$, the Matsubara sum approaches the continuous-frequency integral and Eq.~\eqref{eq:Ftotal} reduces to Eq.~\eqref{eq:Efinal}. The corresponding free-energy and pressure ratios relative to ideal electromagnetism therefore approach
\begin{equation}
R_F(0)=R_P(0)=\frac{3+\sqrt2}{2}\frac{v}{c},
\qquad
R_F(0)=R_P(0)\simeq2.20710678\quad (v=c).
\label{eq:Rlow}
\end{equation}

In the classical limit, $k_BTd/\hbar u\gg1$, the $m\ge1$ Matsubara terms are exponentially suppressed. The $m=0$ contribution of one polarization is independent of $u$,
\begin{equation}
\frac{F_1^{\rm cl}}{A}=-\frac{\zeta(3)k_BT}{16\pi d^2},\qquad
P_1^{\rm cl}=-\frac{\zeta(3)k_BT}{8\pi d^3}.
\end{equation}
All five polarizations therefore contribute equally:
\begin{equation}
\frac{F_{\rm frac}^{\rm cl}}{A}=-\frac{5\zeta(3)k_BT}{16\pi d^2},\qquad
P_{\rm frac}^{\rm cl}=-\frac{5\zeta(3)k_BT}{8\pi d^3}.
\label{eq:classicallimit}
\end{equation}
For ideal electromagnetism with perfectly conducting boundaries, both electromagnetic polarizations contribute in the zero-frequency sector. Hence
\begin{equation}
\frac{F_{\rm EM}^{\rm cl}}{A}=-\frac{\zeta(3)k_BT}{8\pi d^2},\qquad
P_{\rm EM}^{\rm cl}=-\frac{\zeta(3)k_BT}{4\pi d^3},
\label{eq:EMclassic}
\end{equation}
and therefore
\begin{equation}
R_F(\infty)=R_P(\infty)=\frac52.
\label{eq:Rhigh}
\end{equation}
This normalization is the ideal perfect-conductor reference used throughout the paper. It should not be confused with dissipative material prescriptions in which the electromagnetic TE zero-frequency contribution can be absent and the classical reference force is correspondingly different. No such material-response model is used here.

The physical content of the two limits is therefore different: the quantum vacuum weights each polarization by its propagation speed, whereas the classical Matsubara-zero sector counts confined polarizations independently of that speed.

\subsection{Exact pressure and split thermal crossover}

Differentiating Eq.~\eqref{eq:matsubara} with respect to $d$ gives a numerically stable expression for the pressure of one polarization,
\begin{equation}
P_1(d,T;u)=-\frac{k_BT}{\pi}\sideset{}{'}\sum_{m=0}^{\infty}
\left[
\frac{a_m^2}{2d}\operatorname{Li}_1(q_m)
+\frac{a_m}{2d^2}\operatorname{Li}_2(q_m)
+\frac{1}{4d^3}\operatorname{Li}_3(q_m)
\right],
\label{eq:P1exact}
\end{equation}
where $a_m=2\pi mk_BT/(\hbar u)$, $q_m=e^{-2a_md}$ and $\operatorname{Li}_1(q)=-\ln(1-q)$. The $m=0$ limit is finite and reproduces the classical result above. The total pressure is
\begin{equation}
P_{\rm frac}=3P_1(d,T;v)+2P_1(d,T;v/\sqrt2),
\end{equation}
and it satisfies the same scaling identity as the free energy. Here the Casimir pressure denotes the generalized force per unit area conjugate to the plate separation. In an effective condensed-matter realization, where the reflecting layers may be embedded or externally fixed within a host medium, it need not coincide directly with the unconstrained mechanical stress on a freely moving plate. The sign is fixed throughout the ideal identical-plate problem. Equivalently, differentiating the logarithmic kernel in Eq.~\eqref{eq:matsubara} gives
\begin{equation}
P_1(d,T;u)=-2k_BT\sideset{}{'}\sum_{m=0}^{\infty}\int\frac{d^2k_\parallel}{(2\pi)^2}\,
\frac{\kappa_m}{e^{2d\kappa_m}-1}<0,
\label{eq:attractivesign}
\end{equation}
for $T,d,u>0$. Hence varying $T$, $d$, or the positive propagation speed $v$ can change the magnitude and crossover scale but cannot produce repulsion under the boundary conditions considered here.

Introducing the dimensionless thermal variable
\begin{equation}
x=\frac{k_BTd}{\hbar v},
\end{equation}
the slow branch is controlled by $\sqrt2x$ while the fast branch is controlled by $x$. Hence the two branches enter the thermal regime at slightly different scales. Since the separation is only a factor $\sqrt2$, there is no parametrically wide regime in which one branch is deeply classical while the other remains deeply quantum. We therefore refer to this behavior as a \emph{split} or \emph{two-scale thermal crossover}, rather than as three distinct asymptotic regimes.

\begin{figure}[t]
\centering
\includegraphics[width=0.78\textwidth]{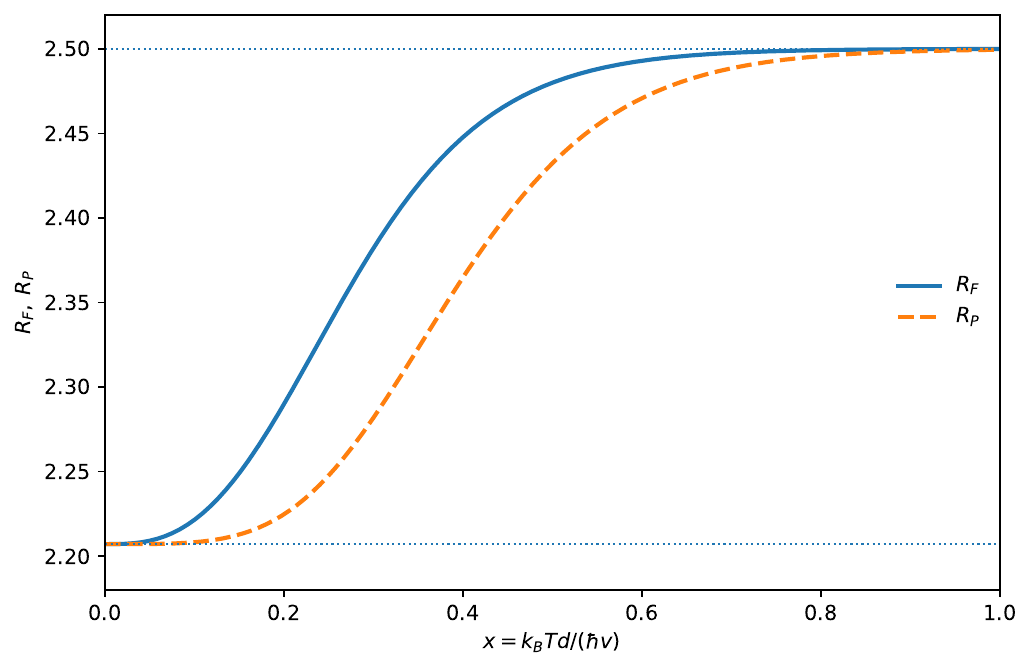}
\caption{Thermal Casimir crossover for $v=c$. The normalized free energy $R_F$ and pressure $R_P$ evolve from the common quantum value $(3+\sqrt2)/2$ toward the classical value $5/2$, with distinct crossover profiles. The slow branch is controlled by $\sqrt2x$, while the fast branch is controlled by $x=k_BTd/(\hbar v)$.}
\label{fig:crossover}
\end{figure}

Figure~\ref{fig:crossover} shows the ratios
\begin{equation}
R_F(x)=\frac{F_{\rm frac}}{F_{\rm EM}},\qquad
R_P(x)=\frac{P_{\rm frac}}{P_{\rm EM}},
\end{equation}
computed from the exact expressions. Both increase numerically from $(3+\sqrt2)/2$ to $5/2$, but the free-energy ratio begins its crossover earlier than the pressure ratio. This difference is explained analytically by the low-temperature expansion below.

\subsection{Low-temperature spectral moments}

For one polarization of speed $u$, expansion of the exact Matsubara expression gives
\begin{equation}
\frac{F_1}{A}=
-\frac{\pi^2\hbar u}{1440d^3}
-\frac{\zeta(3)}{4\pi}\frac{(k_BT)^3}{\hbar^2u^2}
+\frac{\pi^2}{90}d\frac{(k_BT)^4}{\hbar^3u^3}
+\cdots .
\label{eq:lowone}
\end{equation}
Summing the three fast and two slow modes yields
\begin{equation}
\frac{F_{\rm frac}}{A}=
-\frac{(3+\sqrt2)\pi^2\hbar v}{1440d^3}
-\frac{7\zeta(3)}{4\pi}\frac{(k_BT)^3}{\hbar^2v^2}
+\frac{(3+4\sqrt2)\pi^2}{90}d\frac{(k_BT)^4}{\hbar^3v^3}
+\cdots .
\label{eq:lowfracton}
\end{equation}
The successive coefficients involve different moments of the velocity distribution,
\begin{equation}
\sum_a v_a= (3+\sqrt2)v,\qquad
\sum_a v_a^{-2}=\frac7{v^2},\qquad
\sum_a v_a^{-3}=\frac{3+4\sqrt2}{v^3}.
\end{equation}
This is a useful way to characterize what different thermodynamic observables probe in the two-branch spectrum.

A particularly important consequence is that the $T^3$ term in Eq.~\eqref{eq:lowfracton} is independent of $d$ and therefore drops out of the force. Although this term has the same temperature power as a possible surface contribution in a raw Weyl expansion, here it has been obtained \emph{after} the bulk and single-boundary sectors were removed, by expanding the interaction kernel~\eqref{eq:matsubara}; it must therefore not be subtracted again. The same caution applies to the $dT^4$ term in Eq.~\eqref{eq:lowfracton}. Its algebraic form resembles the extensive blackbody contribution, but it is a low-temperature asymptotic term of $F_{\rm int}$, valid only for $k_BTd/(\hbar u)\ll1$. The exact interaction still satisfies $F_{\rm int}\to0$ as $d\to\infty$, so the low-$T$ expansion cannot be used in that limit. The pressure starts receiving a thermal correction only at order $T^4$,
\begin{equation}
P_{\rm frac}=
-\frac{(3+\sqrt2)\pi^2\hbar v}{480d^4}
-\frac{(3+4\sqrt2)\pi^2}{90}\frac{(k_BT)^4}{\hbar^3v^3}
+\cdots .
\label{eq:Plow}
\end{equation}
This explains why $R_F(x)$ departs from its quantum value before $R_P(x)$ in Fig.~\ref{fig:crossover}.

\subsection{Bulk blackbody term and high-temperature corrections}

For an unconstrained branch with speed $u$,
\begin{equation}
\frac{f_{\rm bulk}^{(1)}}{V}=-\frac{\pi^2}{90}\frac{(k_BT)^4}{(\hbar u)^3}.
\end{equation}
The five-mode bulk free-energy density is therefore
\begin{equation}
\frac{f_{\rm bulk}}{V}=-\frac{\pi^2}{90}(3+4\sqrt2)
\frac{(k_BT)^4}{(\hbar v)^3}.
\label{eq:stefanboltzmann}
\end{equation}
This $T^4$ volume term belongs to $F_{\rm bulk}$ in Eq.~\eqref{eq:decomposition} and is removed before the interaction kernel~\eqref{eq:matsubara} is formed. Likewise, any separation-independent surface term generated by the raw mode sum belongs to $F_{\rm self}$ and is not part of the force between the plates. Consequently the high-temperature expansion of the \emph{renormalized interaction} contains neither a surviving $T^4$ blackbody term nor an independent $T^3$ surface term. Once the Matsubara zero mode gives Eq.~\eqref{eq:classicallimit}, all $m\ge1$ corrections are exponentially suppressed, with scales $e^{-4\pi k_BTd/(\hbar v)}$ and $e^{-4\sqrt2\pi k_BTd/(\hbar v)}$ for the fast and slow branches, respectively. Thus the classical Casimir interaction is linear in $T$, as required, while the blackbody and plate self-energy sectors remain absent.

\subsection{Entropy and internal energy}
\label{sec:thermoUS}

From $S=-\partial F/\partial T$ and $U=F+TS$, Eq.~\eqref{eq:lowfracton} gives
\begin{equation}
\frac{S}{A}=\frac{21\zeta(3)}{4\pi}k_B\left(\frac{k_BT}{\hbar v}\right)^2
-\frac{2\pi^2}{45}(3+4\sqrt2)k_Bd\left(\frac{k_BT}{\hbar v}\right)^3+\cdots ,
\label{eq:Slow}
\end{equation}
and
\begin{equation}
\frac{U}{A}=\frac{E_{\rm C}}{A}
+\frac{7\zeta(3)}{2\pi}\frac{(k_BT)^3}{\hbar^2v^2}
-\frac{(3+4\sqrt2)\pi^2}{30}d\frac{(k_BT)^4}{\hbar^3v^3}+\cdots .
\label{eq:Ulow}
\end{equation}
The entropy vanishes as $T^2$, satisfying the Nernst requirement within the ideal model. In the classical limit,
\begin{equation}
\frac{S^{\rm cl}}{A}=\frac{5\zeta(3)k_B}{16\pi d^2},\qquad
\frac{U^{\rm cl}}{A}\to0,
\label{eq:SUhigh}
\end{equation}
up to exponentially small corrections.

\section{Effective Propagation Velocity and Casimir Response}
\label{sec:velocity}

The preceding calculation was deliberately written with the fast-branch speed $v$ explicit. We now use this freedom to separate two logically distinct ingredients of the minimal model. The tensor structure fixes the multiplicities and the relative branch velocity,
\begin{equation}
(v_1,v_2,v_3,v_4,v_5)=\left(v,v,v,\frac{v}{\sqrt2},\frac{v}{\sqrt2}\right),
\label{eq:velocities}
\end{equation}
whereas the common scale $v$ may be interpreted as an effective stiffness parameter. Indeed, before canonical normalization the quadratic action can be parametrized as
\begin{equation}
S_{\chi,\kappa}=\frac12\int dt\,d^3x\,
\left[\chi E_{ij}E^{ij}-\kappa B_{ij}B^{ij}\right],
\qquad
v=\sqrt{\frac{\kappa}{\chi}}.
\label{eq:stiffnessaction}
\end{equation}
The gauge symmetry and the variational boundary conditions are unchanged by positive constant $\chi$ and $\kappa$: the latter multiply the electric and magnetic boundary variations but do not change their homogeneous zeros. After rescaling the field to absorb the overall normalization, Eq.~\eqref{eq:stiffnessaction} reduces to the same minimal tensor theory with an explicit propagation scale. Thus the analysis of the preceding sections is preserved while the absolute frequency scale is varied. Additional independent tensor contractions would define a broader class of quadratic theories and can also modify the ratio $1/\sqrt2$.

It is useful to introduce
\begin{equation}
\eta\equiv\frac{v}{c}.
\label{eq:eta}
\end{equation}
At zero temperature Eq.~\eqref{eq:Rzero} becomes
\begin{equation}
R_P(0;\eta)=R_F(0;\eta)=\frac{3+\sqrt2}{2}\,\eta.
\label{eq:Reta0}
\end{equation}
For $\eta=1$ this gives $2.2071$. The zero-temperature fractonic and electromagnetic pressures have equal magnitude at
\begin{equation}
\eta_*=\frac{2}{3+\sqrt2}\simeq0.4531.
\label{eq:etastar}
\end{equation}
Equation~\eqref{eq:attractivesign} shows that the pressure remains attractive across this comparison.

The classical limit behaves very differently. Since the Matsubara zero mode in Eq.~\eqref{eq:classicallimit} is independent of propagation speed,
\begin{equation}
R_F(\infty;\eta)=R_P(\infty;\eta)=\frac52
\label{eq:Retainfty}
\end{equation}
for every fixed $\eta>0$. Hence thermal evolution erases the dependence on the absolute propagation scale: the quantum ground-state interaction weights the five polarizations by their frequencies, whereas the classical interaction counts them equally.

The intrinsic thermal variable is $x=k_BTd/(\hbar v)$, so the corresponding crossover temperature is parametrically set by
\begin{equation}
T_*\sim\frac{\hbar v}{k_Bd}.
\label{eq:Tstar}
\end{equation}
Its absolute value is therefore strongly material dependent: a reduced emergent gauge-field velocity can bring the thermal crossover to substantially lower temperatures than in the electromagnetic case. Accordingly, decreasing $v$ at fixed $T$ and $d$ moves the fractonic field through the crossover at a lower temperature. Together with Eqs.~\eqref{eq:Reta0} and \eqref{eq:Retainfty}, this completely specifies the velocity dependence: smaller $v$ lowers the quantum endpoint, shifts the crossover to lower temperature, and leaves the classical endpoint $5/2$ unchanged. For $\eta<\eta_*$ the fractonic pressure begins with a smaller magnitude than the electromagnetic pressure and crosses it during the thermal evolution.

The same parameter dependence is visible in the other thermodynamic quantities. In the low-temperature domain $x=k_BTd/(\hbar v)\ll1$, Eqs.~\eqref{eq:lowfracton}, \eqref{eq:Plow}, \eqref{eq:Slow}, and \eqref{eq:Ulow} show the hierarchy
\begin{equation}
E_{\rm C}\propto v,\qquad
\Delta F_{T^3}\propto v^{-2},\qquad
\Delta F_{T^4}\propto v^{-3},\qquad
\Delta P_T\propto v^{-3},
\label{eq:vscalings1}
\end{equation}
and
\begin{equation}
S_{\rm leading}\propto v^{-2},\qquad
\Delta U_{T^3}\propto v^{-2},\qquad
\Delta U_{T^4}\propto v^{-3}.
\label{eq:vscalings2}
\end{equation}
Thus a smaller propagation speed lowers the zero-point scale while making thermal corrections important at a lower temperature. The inverse powers of $v$ in the low-temperature expansion apply only for $x\ll1$; the exact Matsubara expressions remain finite through the crossover and approach classical results in which $F_{\rm frac}^{\rm cl}$, $P_{\rm frac}^{\rm cl}$ and $S^{\rm cl}$ are independent of $v$, while $U^{\rm cl}\to0$.

Here the vacuum is the ground state of the quantized rank-2 gauge field. Its zero-point energy is weighted by $\hbar\omega_a/2$ and therefore retains the effective velocity scale $v$; thermal occupation progressively removes that velocity weighting from the interaction until the classical mode-counting limit is reached.

Since the rank-2 theory is an effective long-wavelength description, its
absolute zero-point energy is cutoff dependent and has no independent
physical significance. The quantity computed here is instead the
separation-dependent interaction obtained after subtracting the bulk and
single-boundary contributions. It is therefore meaningful within the
effective theory provided the plate separation remains large compared with
the microscopic ultraviolet scale at which the continuum rank-2
description breaks down.

\section{Conclusions}
\label{sec:discussion}

We have derived the zero- and finite-temperature Casimir interaction of the
free trace-full scalar-charge rank-2 gauge field between two parallel
reflecting boundaries. The variational boundary problem confines all five
physical polarizations and preserves the two bulk dispersion branches: three
modes propagate with speed $v$ and two with speed $v/\sqrt2$. An explicit
analysis of branch mixing and exceptional momentum sectors shows that no
additional cavity-frequency families are generated by the boundaries. The
interaction spectrum can therefore be organized into five independent
channels once the tensorial boundary problem has been solved.

At zero temperature the ground-state energy weights each polarization by its
frequency. The resulting pressure scales as $d^{-4}$ and, relative to the
ideal electromagnetic result, satisfies
\begin{equation}
\frac{P_{\rm C}}{P_{\rm EM}}
=
\frac{3+\sqrt2}{2}\frac{v}{c}.
\end{equation}
The coefficient directly reflects the velocity-weighted $3+2$ spectrum.
At finite temperature the two branches acquire different thermal weights.
Their characteristic variables, $x$ and $\sqrt2 x$, produce a split crossover
in which the normalized free energy and pressure follow different trajectories
even though they share the same quantum and classical endpoints. The
low-temperature expansion makes the distinction explicit: the $T^3$ term in
the interaction free energy is independent of $d$ and drops out of the
pressure, whose first thermal correction is of order $T^4$.

The high-temperature limit provides a complementary spectral probe. The
Matsubara zero mode is independent of the propagation speed, so the
velocity-weighted quantum result evolves into pure mode counting,
\begin{equation}
R_F,R_P\longrightarrow\frac{5}{2}.
\end{equation}

The calculation defines a benchmark for the minimal free scalar-charge gauge
theory with perfectly reflecting boundaries. More general quadratic tensor
structures can modify the relative branch velocity, and microscopic boundary
dynamics can replace perfect reflection by a frequency- and
polarization-dependent response. In a fractonic medium, thermally activated
dipoles and charges can also screen the higher-rank gauge field
~\cite{pretko2017thermal}. The free-field thermal expressions then apply on
length and time scales preceding effective screening, with the formal $5/2$
endpoint representing the classical limit of the unscreened gauge sector.
These effects provide natural ingredients for a future fractonic Lifshitz
theory based on a microscopic reflection matrix.

The main result is therefore structural: the Casimir interaction converts the
rank-2 polarization content into a directly calculable vacuum and thermal
signature. The variational boundary problem, the $3+2$ two-velocity spectrum,
and the crossover from velocity weighting to mode counting together
distinguish this system from a simple constant rescaling of the electromagnetic
Casimir effect. This provides a self-contained reference point for future
higher-rank gauge realizations in condensed-matter and quantum-simulation
settings.

\section*{Acknowledgments}
CRM acknowledges partial financial support from Conselho Nacional de
Desenvolvimento Cient\'{i}fico e Tecnol\'ogico (CNPq), Grant No.
301122/2025-3. Generative artificial intelligence tools were used during
the preparation of this work for exploratory calculations, consistency
checks, and conceptual refinement. All scientific content, references,
and conclusions were independently verified by the author, who assumes
full responsibility for the manuscript.

\appendix
\section{Plane-wave dynamical matrix and mode check}
\label{app:matrix}

For completeness, we display the algebra underlying the two-branch spectrum. With $\mathbf{k}=(k_x,0,k_z)$, $v=1$ during the algebraic calculation, and the component ordering
\begin{equation}
\mathbf h=(h_{xx},h_{xy},h_{xz},h_{yy},h_{yz},h_{zz})^{T},
\end{equation}
the Fourier-transformed field equation can be written as $M\mathbf h=0$, with
\begin{equation}
M=\begin{pmatrix}
k_z^2-\omega^2 & 0 & -k_xk_z & 0 & 0 & 0\\
0 & k_x^2/2+k_z^2-\omega^2 & 0 & 0 & -k_xk_z/2 & 0\\
-k_xk_z/2 & 0 & (k_x^2+k_z^2)/2-\omega^2 & 0 & 0 & -k_xk_z/2\\
0 & 0 & 0 & k_x^2+k_z^2-\omega^2 & 0 & 0\\
0 & -k_xk_z/2 & 0 & 0 & k_x^2+k_z^2/2-\omega^2 & 0\\
0 & 0 & -k_xk_z & 0 & 0 & k_x^2-\omega^2
\end{pmatrix}.
\label{eq:matrixappendix}
\end{equation}
Its determinant is
\begin{equation}
\det M=-\frac{\omega^2}{4}
(k^2-2\omega^2)^2(k^2-\omega^2)^3,
\qquad k^2=k_x^2+k_z^2,
\end{equation}
reproducing Eq.~\eqref{eq:detM}. Restoring the characteristic speed gives the physical nonzero roots $\omega=v k$ and $\omega=vk/\sqrt2$, with multiplicities three and two, respectively. The remaining $\omega=0$ root is the pure-gauge direction $h_{ij}\propto k_i k_j$, generated by $\delta A_{ij}=\partial_i\partial_j\Lambda$, and is therefore not counted as a propagating polarization.

The scalar-charge Gauss law is not contained merely by choosing $A_\tau=0$; it must still be imposed as
\begin{equation}
k_i k_j h^{ij}=k_x^2h_{xx}+2k_xk_zh_{xz}+k_z^2h_{zz}=0.
\label{eq:gaussappendix}
\end{equation}
Substitution of each of the five polarization vectors in Eq.~\eqref{eq:pols} into Eq.~\eqref{eq:gaussappendix} gives zero. Direct substitution into Eq.~\eqref{eq:matrixappendix} likewise gives $M h^{(a)}=0$ at the corresponding branch frequency. This provides an explicit algebraic check of both the $3+2$ multiplicity and the compatibility of the propagating modes with the Gauss constraint.

\subsection{Exceptional momentum sectors}
\label{app:exceptional}

The polarization basis in Eq.~\eqref{eq:pols} was chosen for generic
$k_x k_z\neq0$ and therefore contains coordinate singularities when one of
the momentum components vanishes. These singularities belong only to the
choice of basis and not to the physical spectrum. For completeness, we
analyze the exceptional sectors directly from
Eq.~\eqref{eq:matrixappendix}.

Consider first $k_x=0$ with $k_z\neq0$. The dynamical matrix becomes
diagonal, while the Gauss constraint reduces to
\begin{equation}
k_z^2 h_{zz}=0,
\end{equation}
so that $h_{zz}=0$. The nonzero-frequency null spaces are then
\begin{align}
\omega^2=k_z^2:
&\qquad
h_{xx},\;h_{xy},\;h_{yy},
\\
\omega^2=\frac{k_z^2}{2}:
&\qquad
h_{xz},\;h_{yz}.
\end{align}
Thus the $3+2$ multiplicity is unchanged. For a plate normal to the
$z$ direction, the essential conditions
$A_{xx}=A_{xy}=A_{yy}=0$ impose Dirichlet parity on the three fast
components. In the slow sector, with $k_x=0$,
\begin{equation}
B^{xz}=-\partial_z A_{yz},
\qquad
B^{yz}=\partial_z A_{xz},
\end{equation}
so the natural magnetic conditions impose Neumann parity. For nonzero
normal momentum both sectors therefore obey
\begin{equation}
k_z=\frac{n\pi}{d},
\qquad n=1,2,\ldots,
\end{equation}
with no additional eigenfrequency. Since the $k_x=0$ sector is of measure zero in the transverse-momentum integral, this analysis serves to establish regularity and completeness of the polarization basis and does not generate an additional contribution to the Casimir interaction.

The complementary axis $k_z=0$, $k_x\neq0$, is also regular when treated
directly. The Gauss constraint now gives
\begin{equation}
h_{xx}=0.
\end{equation}
Before imposing the boundary conditions, the fast branch is spanned by
\begin{equation}
h_{yy},\qquad h_{yz},\qquad h_{zz},
\end{equation}
whereas the slow branch is spanned by
\begin{equation}
h_{xy},\qquad h_{xz}.
\end{equation}
The essential condition $A_{yy}=0$ removes the first fast component,
while the natural condition $E_{zz}=0$ removes $h_{zz}$ for
$\omega\neq0$. The surviving fast mode is therefore $h_{yz}$.
Likewise, $A_{xy}=0$ removes one of the slow modes, leaving $h_{xz}$.
Hence the only physical zero-normal-momentum modes are
\begin{equation}
\omega_{0,f}=v|k_x|,
\qquad
h_{yz}\neq0,
\end{equation}
and
\begin{equation}
\omega_{0,s}=\frac{v|k_x|}{\sqrt2},
\qquad
h_{xz}\neq0.
\end{equation}
Both are independent of the plate separation and therefore belong to the
separation-independent sector removed from the Casimir interaction, in
agreement with Eq.~\eqref{eq:spatialzeromodes}.

It remains to check the threshold values appearing in the branch-mixing
analysis of Sec.~\ref{sec:standing}. Let $p=k_{z,f}$ and
$s=k_{z,s}$ as in Eq.~\eqref{eq:psdefs}. At $p=0$, the fast
zero-normal-momentum mode satisfies the boundary conditions independently
of the slow sector. In the $(h^{(2)},h^{(4)})$ channel, for example, the
fast contribution has constant $A_{yz}$ and vanishing $A_{xy}$, while the
slow contribution must still satisfy separately
\begin{equation}
A_{xy}=0,
\qquad
\partial_z A_{yz}=0,
\end{equation}
at both plates, giving $\sin(sd)=0$. No cancellation with the constant
fast mode can relax this condition. The same conclusion holds in the
$(h^{(3)},h^{(5)})$ sector: the essential condition on $A_{xx}$ fixes the
slow amplitude independently, while $E_{zz}=0$ eliminates the potentially
mixing fast $A_{zz}$ component.

At the complementary threshold $s=0$, the regular slow mode that survives
the boundary conditions is precisely the constant $h_{xz}$ mode identified
above. For the same frequency the fast normal wave number is imaginary,
so the fast solution is evanescent rather than a propagating cavity mode.
The conditions $A_{xy}=0$ and $B^{xz}=0$, or respectively
$A_{xx}=0$ and $E_{zz}=0$, form a nonsingular pair for that evanescent
fast solution and set its amplitude to zero. Thus the $s=0$ threshold
also introduces no additional mixed eigenfrequency.

Consequently, the apparent singularities of the generic polarization
vectors and of the intermediate mixing matrices are only coordinate
artifacts. The exceptional momentum sectors preserve the same five
physical degrees of freedom, introduce no additional fast--slow hybrid
families, and leave the interaction-relevant quantization
$k_z=n\pi/d$ unchanged.

\bibliographystyle{JHEP}
\bibliography{fracton_casimir_JHEP_clean}

\end{document}